\documentclass[aip,jcp,preprint,amsmath,amssymb]{revtex4-1}
\usepackage[T1]{fontenc}
\usepackage[utf8]{inputenc}
\usepackage{tipa}
\usepackage{lmodern}
\usepackage{natbib}
\usepackage{dcolumn}
\usepackage[version=3]{mhchem}
\usepackage{graphicx}
\usepackage{color}
\usepackage[normalem]{ulem}
\usepackage{paralist}

\newcolumntype{d}[1]{D{.}{.}{#1}}
\newcommand{\rhos}{\rho_{\sigma}}
\newcommand{\rhoa}{\rho_{\alpha}}
\newcommand{\rhob}{\rho_{\beta}}
\newcommand{\taua}{\tau_\alpha}
\newcommand{\rhosp}{\rho_{\sigma'}}
\newcommand{\rs}{r_\mathrm{s}} 
\newcommand{\rsaa}{r_\mathrm{s}^{\alpha\alpha}} 
\newcommand{\rsab}{r_\mathrm{s}^{\alpha\beta}} 

\newcommand{\pair}{P_{2\lambda}^{\sigma\sigma'}}
\newcommand{\hxcab}{h_{\mathrm{XC}\lambda}^{\alpha\beta}}
\newcommand{\hxcaa}{h_{\mathrm{XC}\lambda}^{\alpha\alpha}}
\newcommand{\hxss}{h_\mathrm{X}^{\sigma\sigma'}}
\newcommand{\hcss}{h_{\mathrm{C}\lambda}^{\sigma\sigma'}}
\newcommand{\hcaa}{h_{\mathrm{C}\lambda}^{\alpha\alpha}}
\newcommand{\hcab}{h_{\mathrm{C}\lambda}^{\alpha\beta}}
\newcommand{\hxaa}{h_{\mathrm{X}}^{\alpha\alpha}}
\newcommand{\bab}{B_{\alpha\beta}}
\newcommand{\baa}{B_{\alpha\alpha}}
\newcommand{\aopp}{a_{\alpha\beta}}
\newcommand{\Aopp}{\mathcal{A}_{\alpha\beta}}
\newcommand{\bopp}{b_{\alpha\beta}}
\newcommand{\Bopp}{\mathcal{B}_{\alpha\beta}}
\newcommand{\copp}{c_{\alpha\beta}}
\newcommand{\dopp}{d_{\alpha\beta}}
\newcommand{\apar}{a_{\alpha\alpha}}
\newcommand{\Apar}{\mathcal{A}_{\alpha\alpha}}
\newcommand{\bpar}{b_{\alpha\alpha}}
\newcommand{\Bpar}{\mathcal{B}_{\alpha\alpha}}
\newcommand{\cpar}{c_{\alpha\alpha}}
\newcommand{\dpar}{d_{\alpha\alpha}}
\newcommand{\hxcss}{h_{\mathrm{XC}\lambda}^{\sigma\sigma'}}
\newcommand{\vext}{\hat{v}_{\mathrm{ext}}}
\newcommand{\rij}{r_{\mathrm{ij}}}

\newcommand{\veeop}{\hat{V}_\mathrm{ee}}

\newcommand{\vca}{V_\mathrm{C}^\lambda}
\newcommand{\vcab}{V_\mathrm{C}^{\alpha\beta,\lambda}}
\newcommand{\vcaa}{V_\mathrm{C}^{\alpha\alpha,\lambda}}
\newcommand{\vcss}{V_\mathrm{C}^{\sigma\sigma',\lambda}}

\newcommand{\kinop}{\hat{T}}
\newcommand{\kins}{T_\mathrm{s}}
\newcommand{\ucoul}{U}
\newcommand{\exc}{E_{\mathrm{XC}}}
\newcommand{\ec}{E_\mathrm{C}}
\newcommand{\ecab}{E_\mathrm{C}^{\alpha\beta}}
\newcommand{\ecba}{E_\mathrm{C}^{\beta\alpha}}
\newcommand{\ecaa}{E_\mathrm{C}^{\alpha\alpha}}
\newcommand{\ecbb}{E_\mathrm{C}^{\beta\beta}}

\newcommand{\diff}{\mathrm{d}}
\newcommand{\rhodep}{\left[\rho\right]}
\newcommand{\psimin}{\Psi^\mathrm{min}_\rho}
\newcommand{\phimin}{\Phi^\mathrm{min}_\rho}
\newcommand{\psimina}{\Psi^\mathrm{min,\lambda}_\rho}
\newcommand{\ra}{\mathbf{r}_1}
\newcommand{\rb}{\mathbf{r}_2}
\newcommand{\rc}{\mathbf{r}_3}
\newcommand{\rn}{\mathbf{r}_N}
\newcommand{\ri}{\mathbf{r}_i}
\newcommand{\dra}{\diff^3\ra}
\newcommand{\drb}{\diff^3\rb}
\newcommand{\braket}[2]{\left\langle #1\middle|#2\middle|#1\right\rangle} 

\newcommand{\refr}[1]{Eq.~(\ref{#1})}

\begin{document}

\title{A first-principles-based correlation functional for harmonious
connection of short-range correlation and long-range dispersion}

\author{Marcin Modrzejewski}
\email{modrzej@tiger.chem.uw.edu.pl}
\affiliation{Faculty of Chemistry, University of Warsaw, 02-093 Warsaw, Pasteura 1, Poland}

\author{Michał Lesiuk}
\affiliation{Faculty of Chemistry, University of Warsaw, 02-093 Warsaw, Pasteura 1, Poland}

\author{Łukasz Rajchel}
\affiliation{Interdisciplinary Centre for Mathematical and Computational Modelling, University of Warsaw, 02-093 Warsaw, Pawińskiego 5a, Poland}

\author{Małgorzata M. Szczęśniak}
\affiliation{Department of Chemistry, Oakland University, Rochester,
  Michigan 48309-4477, USA}

\author{Grzegorz Chałasiński}
\affiliation{Faculty of Chemistry, University of Warsaw, 02-093 Warsaw, Pasteura 1, Poland}

\begin{abstract}
  We present a physically motivated correlation functional belonging to the 
  meta-generalized gradient approximation~(meta-GGA) rung, which can be supplemented
  with long-range dispersion corrections without introducing double-counting of
  correlation contributions. The functional is derived by the method of constraint satisfaction, starting
  from an analytical expression for a real-space spin-resolved correlation hole. The model contains
  a position-dependent function that
  controls the range of the interelectronic correlations described by the semilocal functional. With minimal 
  empiricism, this function may be adjusted so that the correlation model blends with
  a specific dispersion correction describing long-range
  contributions. For a preliminary assessment, our functional has been combined with the DFT-D3 dispersion
  correction and full Hartree-Fock (HF)-like exchange. Despite the HF-exchange approximation,
  its predictions compare favorably with reference interaction energies in an extensive set of non-covalently bound dimers.
\end{abstract}

\maketitle

\section{Introduction}
Inclusion of the dispersion interactions into the set of phenomena accounted for
by DFT models is recognized as one of the challenges in the
development of new density functional approximations~(DFAs).\cite{dobson2012dispersion,dobson2005soft,cohen2012challenges,dobson2012calculation}
 Several
ways have been proposed to correct the currently available semilocal (SL) DFAs for the lacking nonlocal (NL) correlation contribution responsible for the dispersion interactions.\cite{cohen2012challenges,dobson2012dispersion}
Hereafter, global hybrid and range-separated
hybrid functionals will be called SL DFAs. Although the exchange parts of such functionals
are nonlocal, our focus will be on the correlation contributions, which in this
case depend on variables calculated at a single point of space.
The examples of such dispersion-corrected methods are:
\begin{inparaenum}[(i)]
 \item the exchange-hole dipole method
   (XDM),\cite{becke2005exchange,becke2005density,becke2007unified,hesselmann2009derivation,angyan2007exchange}
 \item the atom pairwise additive schemes of Goerigk and Grimme, DFT-D3,\cite{grimme2010consistent} and Tkatchenko-Scheffler approach,\cite{tkatchenko2009accurate}
 \item seamless van der Waals density functionals.\cite{dion2004van,lee2010higher,vydrov2009improving,vydrov2009nonlocal,vydrov2010nonlocal,vydrov2012nonlocal,dobson2012calculation}
 \end{inparaenum}
It is clear that the accuracy of these methods depends not only on
a faithful representation of long-range electronic correlations, but also on
a consistent matching of a dispersion correction and the chosen SL complement. 

Several groups have studied the conditions under which an SL functional
can be incorporated into a dispersion-corrected method.\cite{pernal2009dispersionless,rajchel2010derivation,kamiya2002density,murray2009investigation}
 It has been concluded that the improper behavior of a GGA exchange functional in
 the density tail (large reduced gradient regime) is responsible for
artificial exchange binding (as for the PBE\cite{perdew1996generalized} exchange) or overly repulsive
interaction (as for the B88\cite{becke1988density} exchange).\cite{kamiya2002density,murray2009investigation} Such
systematic errors may cause the NL correction to worsen the results
compared to the bare SL functional. The following exchange
functionals: PW86,\cite{perdew1986accurate}
refitted PW86,\cite{murray2009investigation} and range-separated hybrids\cite{kamiya2002density,vydrov2010nonlocal} were found to be free
from artificial binding, thus being consistent with NL dispersion
correction. Similarly, combining exact
exchange with NL correlation performs satisfactorily.\cite{vydrov2010implementation}

It has been observed that a failure to
satisfy the condition of vanishing correlation for a rapidly varying
density by SL correlation functionals\cite{perdew1996generalized} leads to a systematic overbinding of non-covalent complexes.\cite{kamiya2002density} The \emph{ad hoc} cure is to cancel the error of the correlation by the opposite-sign error of an exchange component.\cite{kamiya2002density} However, this does not resolve the problem of the double counting of SL and NL correlation. Several remedies have been proposed. For atom pairwise schemes, multiple damping functions have been devised.\cite{grimme2011effect} For VV09\cite{vydrov2009nonlocal}
and VV10\cite{vydrov2010nonlocal} density functionals the problem is avoided by demanding the NL constituent to vanish in
the homogeneous electron gas (HEG) limit, because the SL constituent is able
 to describe the whole range of the electronic interactions in this limit.
Finally, \citet{pernal2009dispersionless} devised a procedure to reoptimize an existing SL
exchange-correlation functional so as to recover the dispersionless interaction energy.\cite{pernal2009dispersionless} The rationale of such an approach
is to let the SL functional contribute only the terms that it can describe reliably.

At this point we would like to shed light on the dispersion problem\cite{dobson2002prediction} in DFT. It has been well established that SL functionals fail to recover the long-range multipole-expanded dispersion energy. In fact, nearly all dispersion-corrected DFT approaches aim at recovering only long-range dispersion, roughly determined by
the leading terms of the
multipole expansion: $C_6$, and possibly $C_8$. The exceptions are 
the approaches of \citet{pernal2009dispersionless} and
and \citet{rajchel2010derivation} which supplement an SL functional
 with total non-expanded dispersion from SAPT. It is often overlooked that the long-range contribution, however, does not constitute the whole
dispersion energy at near-equilibrium distances. Setting aside the exchange-dispersion part, the dispersion energy,
as defined in the symmetry-adapted perturbation theory~(SAPT),\cite{hesselmann2003intermolecular,misquitta2005intermolecular} has a complex nature, and includes both long- and short-range contributions. This has first been observed by \citet{koide1976new} who quantified the short-distance behavior of the dispersion energy as $A + B R^2$, with $R$ being the intermonomer separation. The importance of the short-range correlation is also unambiguously, though indirectly, supported
 by the significance of bond functions and explicitly
 correlated Gaussian geminals in the dispersion energy calculations.\cite{burcl1995role}
 A more direct argument points to 
 the existence of short-range terms in the exact angular expansion
 of the dispersion energy.  In the case of atomic interactions, the latter involves the interaction between $S$ states of monomers,\cite{koide1976new} which give no
contribution to the multipole expansion. Numerical results show that near the equilibrium bond length these terms, decaying
 exponentially with the overlap density, can be comparable in magnitude to the 
 multipole expansion terms.\cite{koide1981second} As demonstrated by~\citet{dobson2002prediction} with the aid of simple models,
 SL functionals cannot recover the
 multipole expansion of the dispersion
 energy. However, there is a good reason to believe that SL functionals are capable of describing
 the terms that depend on overlap density.\cite{dobson2002prediction} Our model is intended to capture this contribution.
 
Recent thorough assessments of
DFT-D, XDM, and VV10 approaches have clearly shown
that a combination of an SL functional specifically designed for
a dispersion-corrected treatment with a dispersion correction improves both the description of
noncovalent
interactions\cite{burns2011density,goerigk2011thorough,hujo2011performance,vazquez2010assessment,thanthiriwatte2011assessment} and
general molecular properties.\cite{goerigk2011thorough,hujo2011performance} Among the functionals
that use atom-pairwise DFT-D correction,
B97-D,\cite{grimme2006semiempirical}
B97-D3,\cite{goerigk2011thorough}
 and $\omega$-B97X-D\cite{chai2008long,chai2008systematic}
are characterized by one of the smallest magnitude and spread of
errors in interaction energies\cite{burns2011density} while performing
well in thermochemistry and reaction kinetics. The methods utilizing unaltered
conventional SL functional suffer from systematic errors. For example, PBE0-D2,
PBE-D3, and B3LYP-D3 tend to underbind dispersion-bound complexes and overbind
hydrogen-bonded systems.\cite{burns2011density}
The systematic overbinding of charge-transfer complexes within DFT is more pronounced for the dispersion-corrected approaches than for the pure DFAs.\cite{ruiz1996charge,sini2011evaluating}
 See~Ref.\citenum{sini2011evaluating} and Table~2
 in Ref.~\citenum{podeszwa2010extension}, where numerical
examples of huge overbinding by $\omega$-B97X-D and
B97-D functionals applied to charge-transfer complexes are given.

Although much attention has been devoted to the development of the 
proper exchange contribution,\cite{kamiya2002density,murray2009investigation,swart2009new} 
the theoretical effort to derive a dispersion-consistent SL correlation
functional thus far has been reduced to reoptimizing known expressions.
It has also been observed\cite{grimme2006semiempirical,chai2008systematic}
 that fitting to empirical data coupled with
addition of higher-order terms in the B97 expansion\cite{becke1997density} does not systematically improve the performance as the saturation is approached. Clearly, there is a demand for the theoretical effort to overcome the problem.

The aim of this work is to develop an SL correlation functional
that can be matched with an arbitrary long-range dispersion correction
by optimizing a single parameter that has a simple physical meaning.
 As a demonstration of this approach, we will combine our approximation with
 DFT-D3 dispersion correction,\cite{grimme2010consistent} which contributes damped $C_6/R^6 + C_8/R^8$ terms, with
 no short-range contributions. To avoid the systematic error of
 spurious exchange attraction, full HF exchange will be used. To match SL and NL constituents,
 a function which controls the spatial extent of our SL correlation hole will be
 adjusted by minimization of errors in a relevant set of molecules.
 
\section{Theory}

We consider an electronic ground state of a finite
 molecular system described by an electronic Hamiltonian of the form
\begin{equation}
\label{hamiltonian}
\hat{H}= \kinop + \sum_i \vext(\ri) + \veeop,
\end{equation}
where $\kinop$ is the kinetic energy operator, the multiplicative external potential $\vext$ is taken to be the
 Coulomb potential of nuclear attraction, and $\veeop$ is the interelectronic repulsion. Atomic units are 
assumed throughout this work. In constrained-search formulation\cite{levy1979universal}
 of DFT\cite{hohenberg1964inhomogeneous} the ground state energy of electronic
 system can be expressed as
\begin{equation}
\label{constrained-search}
E_0=\min_\rho \left[ \int \vext(\ra)\rho(\ra)\diff^3 \ra + \braket{\psimin}{\kinop+\veeop}  \right],
\end{equation}
where $\psimin$ denotes an $N$-body wavefunction that yields electronic
 density $\rho$ and simultaneously minimizes expectation
value of $\kinop+\veeop$. In Kohn-Sham scheme\cite{kohn1965self} the second
term on the right-hand side of \refr{constrained-search} is decomposed
 into noninteracting kinetic, Hartree,
 and exchange-correlation energies, respectively:
\begin{equation}
\braket{\psimin}{\kinop+\veeop}=\kins\rhodep+\ucoul\rhodep+\exc\rhodep.
\end{equation}
Noninteracting kinetic energy $\kins$ is known explicitly in terms of
 the wavefunction of the KS system, denoted here as $\phimin$,
which merely minimizes the expectation value of $\kinop$:
\begin{equation}
\kins\rhodep=\braket{\phimin}{\kinop}.
\end{equation}
Hartree energy is given by a classical formula
\begin{equation}
\ucoul\rhodep=\frac{1}{2}\iint \frac{\rho(\ra)\rho(\rb)}{r_{12}} \dra\drb.
\end{equation}
Exchange-correlation energy can be formally expressed through adiabatic connection
formula\cite{levy1996elementary}
\begin{equation}
\label{adiabatic-connection}
\exc\rhodep=\int_0^1 \braket{\psimina}{\veeop}\diff\lambda -\ucoul\rhodep,
\end{equation}
where $\psimina$ minimizes the expectation value of $\kinop+\lambda\veeop$ and yields
the same electronic density as wavefunction at $\lambda=1$. 
\refr{adiabatic-connection} can be further decomposed so that the correlation energy
 is separately expressed through coupling-constant integral
\begin{equation}
\ec\rhodep=\int_0^1\vca\rhodep \diff\lambda \label{lambda-average}
\end{equation}
where 
\begin{equation}
\vca\rhodep = \braket{\psimina}{\veeop} - \braket{\phimin}{\veeop}. \label{vc-def}
\end{equation}
Approximating $\vca$ is the primary objective of this work. Let us begin by expressing
 $\vca$ in terms of a $\lambda$-dependent correlation hole
\begin{equation}
\vca=\frac{1}{2}\sum_{\sigma\sigma'}\iint \frac{\rho_\sigma(\ra)\hcss(\ra,\rb)}{r_{12}}\dra \drb,
\end{equation}
where $\sigma$ denotes a spin variable,
\begin{equation}
\hcss(\ra,\rb) = \hxcss(\ra,\rb) - \hxss(\ra,\rb) \label{correlation-hole-def}
\end{equation}
and
\begin{align}
\hxcss(\ra,\rb)&=\frac{\pair(\ra,\rb)}{\rhos(\ra)}-\rhosp(\rb), \\
\hxss(\ra,\rb)&= -\delta_{\sigma\sigma'} \frac{\left|\sum_i^{N_\sigma}\psi_{i\sigma}^*(\ra) \psi_{i\sigma}(\rb)\right|^2}{\rho_\sigma(\ra)}.
\end{align}
$N_\sigma$ is a number of $\sigma$-spin electrons and pair probability density, $\pair(\ra,\rb)$, is defined as 
\begin{align}
\pair(\ra,\rb)&=N(N-1)  \nonumber \\
& \times \sum_{\sigma_3\cdots\sigma_N}\int \Psi_\rho^{\min,\lambda*}(\ra\sigma,\rb\sigma',\ldots,\rn\sigma_N)
\nonumber \\
& \times
\Psi_\rho^{\min,\lambda}(\ra\sigma,\rb\sigma',\ldots,\rn\sigma_N)
\nonumber \\
& \times \diff^3 \rc \cdots \diff^3 \rn.
\end{align} 
Note that, due to the symmetry of $\rij^{-1}$ operator, it is the spherical average of exchange-correlation hole around
 the reference electron that enters the energy expression:
\begin{equation}
\begin{split}
\vcss &= \frac 1 2 \iint \frac{\rho_\sigma(\ra)\hcss(\ra,\rb)}{r_{12}}\dra \drb \\
&= \frac 1 2 \int \dra \int_0^\infty \frac{\rho_\sigma(\ra)\hcss(\ra,s)}{s} 4\pi s^2 \diff s \label{vcss-spherical}
\end{split}
\end{equation}
where the spherical average is implied by scalar argument $s$,
\begin{equation}
\hcss(\ra, s) = \frac{1}{4\pi} \int_0^{2\pi} \diff \phi_\mathbf{s} \int_0^\pi \hcss(\ra, \ra + \mathbf{s})
 \sin \theta_\mathbf{s} \diff \theta_\mathbf{s}. \label{spherical-average}
\end{equation}
\refr{vcss-spherical} means that without loss of generality
 we can focus our attention on approximating isotropic quantity defined in \refr{spherical-average}.

We postulate the following form of opposite-spin and same-spin correlation holes:
\begin{align}
  \hcab(\ra,s) &= (\aopp + \bopp s + \copp s^2) \exp(-\dopp s), \label{opphole-def}\\
  \hcaa(\ra,s) &= s^2 (\apar + \bpar s + \cpar s^2) \exp(-\dpar s). \label{parhole-def}
\end{align}
Unknown parameters appearing above 
are actually functions of $\rho(\ra)$, but we will not write that explicitly for the sake of brevity. Quadratic behavior of $\hcaa$ near the reference point
results from the Pauli exclusion principle and the cusp condition
discussed further below. The model of \refr{opphole-def} and \refr{parhole-def} cannot
recover radial dependence of the true hole at each point,
 however, its simple form leads to reasonable shape of the
system-averaged correlation hole.  We thus assume that the wealth of 
features of the exact hole\cite{baerends1997quantum} is
averaged out, and use system-averaged function of shape similar to
\refr{opphole-def} and \refr{parhole-def} in the energy
expression. The argument involving system average was originally put forward
 by Burke~et~al.\cite{burke1998semilocal} in
their discussion of the success of the local-density approximation~(LDA).

\begin{figure}[p]
\includegraphics[width=1.0\textwidth]{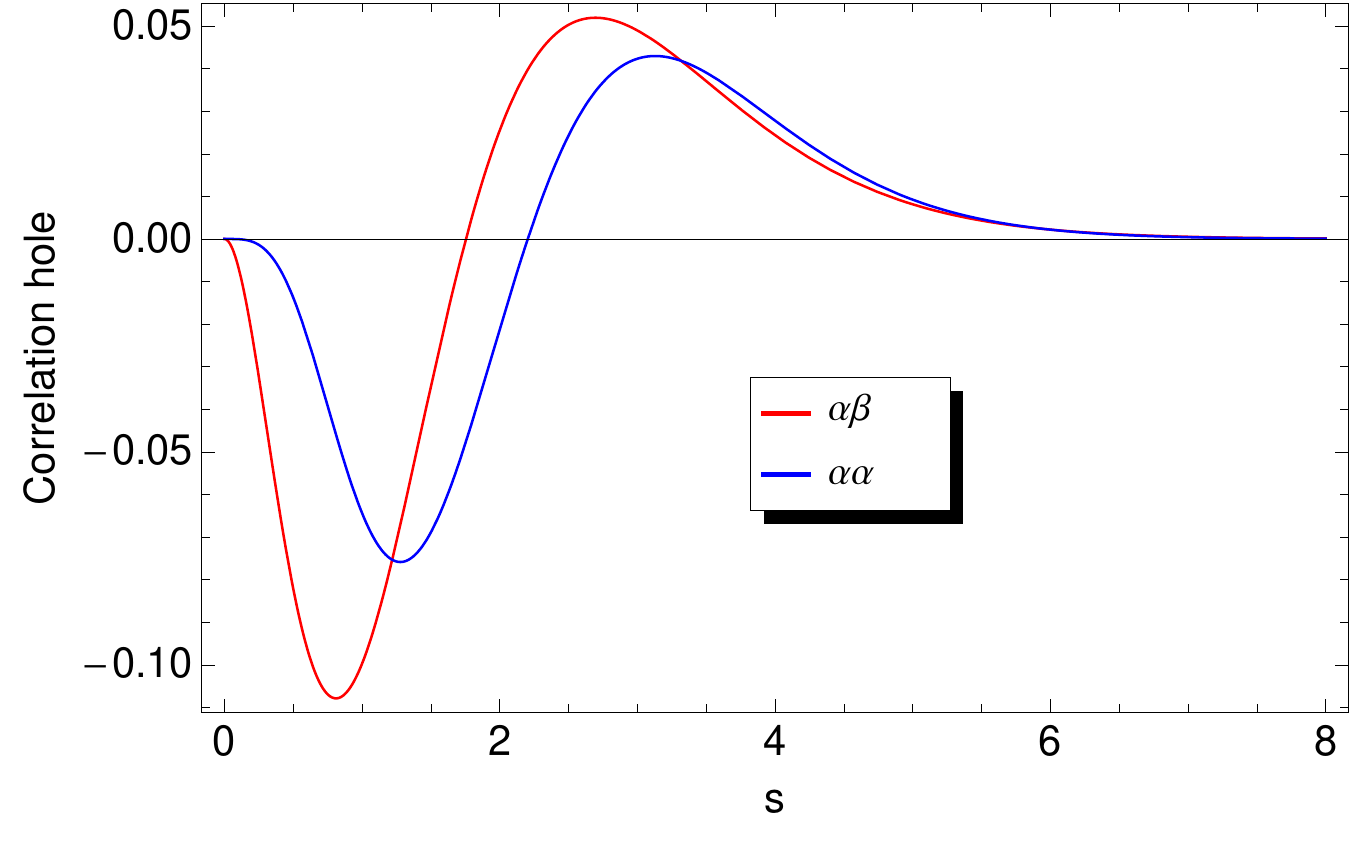}
\caption{Shape of the approximate correlation holes defined in
  Eqs~\ref{opphole-def} and~\ref{parhole-def}. The functions
  illustrated on the graph are $h_\mathrm{C}^{\alpha\beta,1}$
  and $h_\mathrm{C}^{\alpha\alpha,1}$
  multiplied by $4 \pi s^2$. The functions are
  evaluated at $\rs=1$ in the spin-compensated HEG limit. Atomic units are used.}
\label{fig-corrhole}
\end{figure}

The shape of the correlation holes corresponding to Eqs.~\ref{opphole-def}
and~\ref{parhole-def} is illustrated
in~Fig.~\ref{fig-corrhole} (the details of parametrization are discussed below.) For any spin density
the qualitative picture is similar: both same-spin and opposite-spin
holes are removing electrons in the vicinity of the origin,
then cross the abscissa exactly once, and decay
exponentially. The fact that both model correlation holes~\eqref{opphole-def}--\eqref{parhole-def} change sign
exactly once can be readily proven.

The form of correlation holes given in~\refr{opphole-def} and~\refr{parhole-def}
has been derived from observation of system-averaged correlation holes
(correlation intracules) in simple systems dominated by dynamical
correlation. Qualitatively, the shape of our model correlation hole is
similar to correlation intracules in \ce{He},\cite{sarsa1998correlated} 
\ce{Ne},\cite{sarsa1998correlated} and \ce{H2}\cite{hollett2011nature} near
the equilibrium bond length. We note that there is a qualitative discrepancy
between our approximate correlation hole and the accurate one for systems like \ce{Li}\cite{sarsa1998correlated} or
 \ce{Be}.\cite{sarsa1998correlated} These systems are characterized by a
 significant contribution of static correlation. However, there
 is a substantial cancellation between exchange and correlation holes in
 systems of this type.

The cusp conditions for same-spin and opposite-spin
exchange-correlation holes\cite{rajagopal1978short} considerably
restrict the short-range expansion\cite{becke1988correlation} of $\hxcab$ and $\hxcaa$:
\begin{align}
  \hxcab(\ra,s) &= (\bab - \rhob) + \lambda  \bab  s + \ldots, \label{hxcab-expansion}\\
  \hxcaa(\ra,s) &= -\rhoa + (\baa - \frac{1}{6} \nabla^2 \rhoa)s^2 + \frac{\lambda}{2}\baa s^3 + \ldots \label{hxcaa-expansion}
\end{align}
\refr{correlation-hole-def} together with \refr{hxcab-expansion}, \refr{hxcaa-expansion}, and the expansion of
spherically-averaged exact exchange hole valid at zero current density,\cite{becke1988correlation,lee1987gaussian,becke1996current}
\begin{equation}
\hxaa(\ra,s) = -\rhoa - \frac{1}{6} \left[ \nabla^2\rhoa -2 \taua + \frac{1}{2} \frac{(\nabla\rhoa)^2}{\rhoa} \right]s^2+\ldots, \label{exchange-hole}
\end{equation}
yields short-range expansions of correlation holes:\cite{becke1988correlation}
\begin{align}
  \hcab(\ra,s) &= (\bab - \rhob) + \lambda  \bab  s + \ldots, \label{hcab-expansion}\\
  \hcaa(\ra,s) &= \left( \baa - \frac 1 3 D_\alpha \right)s^2 + \frac \lambda 2 \baa s^3 + \ldots \label{hcaa-expansion}
\end{align}
$D_\alpha$ is always non-negative and vanishes for single orbital densities:
\begin{equation}
D_\alpha = \taua - \frac{|\nabla\rho_\alpha|^2}{4\rhoa}, \label{d-inhom}
\end{equation}
where $\taua$ is essentially the density of noninteracting kinetic energy
\begin{equation}
\taua = \sum_i^{N_\alpha} |\nabla\psi_{i\alpha}|^2.
\end{equation}
We will adjust the unknown functions $\bab$ and $\baa$ in \refr{hxcab-expansion} and \refr{hxcaa-expansion}
to recover short range expansion of spin-resolved pair distribution
 function of the HEG developed by
 Gori-Giorgi and Perdew.\cite{gori2001short} The pair distribution
 function represents the
 solution of the Overhauser model.\cite{overhauser1995pair}
 $\baa$ will be further modified to eliminate self-interaction error of the correlation functional.
We leave the on-top value of the correlation hole
(determined solely by $\bab$) unchanged by inhomogeneity corrections
because it is well-transferable from the HEG to real
systems.\cite{burke1998semilocal} For the discussion of the quality of
the HEG on-top hole density see the work of \citet{burke1998semilocal}.

Comparison of homogeneous density limit of \refr{hxcab-expansion} and \refr{hxcaa-expansion} with
short-range expansions of the spin-resolved HEG pair distribution
function\cite{gori2001short} yields
\begin{align}
\begin{split}
\bab(\rhoa, \rhob,\lambda) &= \rhob \left(1 + 0.0207\lambda\rsab + 0.08193(\lambda\rsab)^2 \right. \\ 
&\left. -0.01277(\lambda\rsab)^3+0.001859(\lambda\rsab)^4 \right)\exp(-0.7524\lambda \rsab),
\end{split} \label{bab-heg}
\\
\begin{split}
\baa^\mathrm{HEG}(\rhoa,\lambda) &= \frac{D_\alpha^\mathrm{HEG}}{3} \left(1 -
0.01624\lambda\rsaa + 0.00264(\lambda\rsaa)^2 \right) \\
& \times \exp(-0.5566\lambda\rsaa), \label{baa-heg}
\end{split}
\end{align}
where $\rsaa$ and $\rsab$ introduce the dependence on electronic spin densities,
\begin{align}
  \rsaa &= \frac{\left( 3/\pi \right)^{1/3}}{2
    \rhoa^{1/3}} \label{rsaa-def} \\
  \rsab &= \frac{\left( 3/\pi \right)^{1/3}}{\rhoa^{1/3} +
    \rhob^{1/3}} \label{rsab-def}
\end{align}
and each of them reduces to the Seitz radius
\begin{equation}
  \rs = \left( \frac{3}{4\pi\rho} \right)^{1/3},
\end{equation}
for spin-compensated systems. The formulae~\eqref{bab-heg} and~\eqref{baa-heg} respect the exact high-density
expansion derived by \citet{rassolov2000reply}. The HEG limit of parameter~\eqref{d-inhom} in \eqref{baa-heg} reads\cite{becke1988correlation}
\begin{equation}
D^\mathrm{HEG}_\alpha=\frac{3}{5} (6\pi^2)^{2/3} \rho_\alpha^{5/3}.
\end{equation}
We substitute $D^\mathrm{HEG}_\alpha$ in \refr{baa-heg} for $D_\alpha$
of \refr{d-inhom} to get $\baa$:
\begin{equation}
\begin{split}
\baa(\rho_\alpha,|\nabla\rho_\alpha|,\tau_\alpha,\lambda) &= \frac{D_\alpha}{3} \left(1 -
0.01624\lambda\rsaa + 0.00264(\lambda\rsaa)^2 \right) \\
& \times \exp(-0.5566\lambda\rsaa). \label{baa-inhom}
\end{split}
\end{equation}
Such choice of the $\baa$ function leads to vanishing parallel spin correlation
contribution for single orbital densities. In that sense no correlation
 self-interaction error is present.

Restricting undetermined coefficients in \refr{opphole-def}
and \refr{parhole-def} to yield short-range expansions of \refr{hcab-expansion}
and \refr{hcaa-expansion}, respectively, gives
\begin{align}
\aopp &= \bab - \rho_\beta, \\
\bopp &= \lambda \bab + \dopp \aopp. \label{bopp-param}
\end{align}
Correct shape of $\hcab$ can be ensured requiring that the function satisfies the appropriate
 sum rule,
\begin{equation}
4\pi\int_0^\infty \hcab(\ra,s)s^2\diff s = 0. \label{sum-rule}
\end{equation}
Consequently, coefficient $\copp$ is fixed for \refr{sum-rule} to hold for all densities:
\begin{equation}
\copp = -\frac{1}{12}(\aopp\dopp^2+3\bopp\dopp) \label{copp-param}.
\end{equation}
Analogously,
\begin{align}
\apar &= \baa - \frac{D_\alpha}{3}, \\
\bpar &= \frac{\lambda}{2}\baa + \apar\dpar, \label{bpar-param} \\
\cpar &= -\frac{1}{30}(\apar\dpar^2 + 5\bpar\dpar) \label{cpar-param}.
\end{align}
With all but $\dopp$ and $\dpar$ coefficients determined, spin resolved contributions
to $\vca$,
\begin{equation}
\vcss = \frac 1 2 \int \diff^3 \ra \int_0^\infty \frac{\rhos \hcss(\ra,s)}{s} 4\pi s^2 \diff s,
\end{equation}
can now be given as:
\begin{align}
\vcab &= \int \diff^3 \ra \rhoa \pi \frac{\bopp +
  \aopp\dopp}{\dopp^3}, \label{vcabl} \\
\vcaa &= \int \diff^3 \ra \rhoa \pi \frac{8\bpar +
  4\apar\dpar}{\dpar^5}. \label{vcaal}
\end{align}

\begin{figure}[p]
\includegraphics[width=1.0\textwidth]{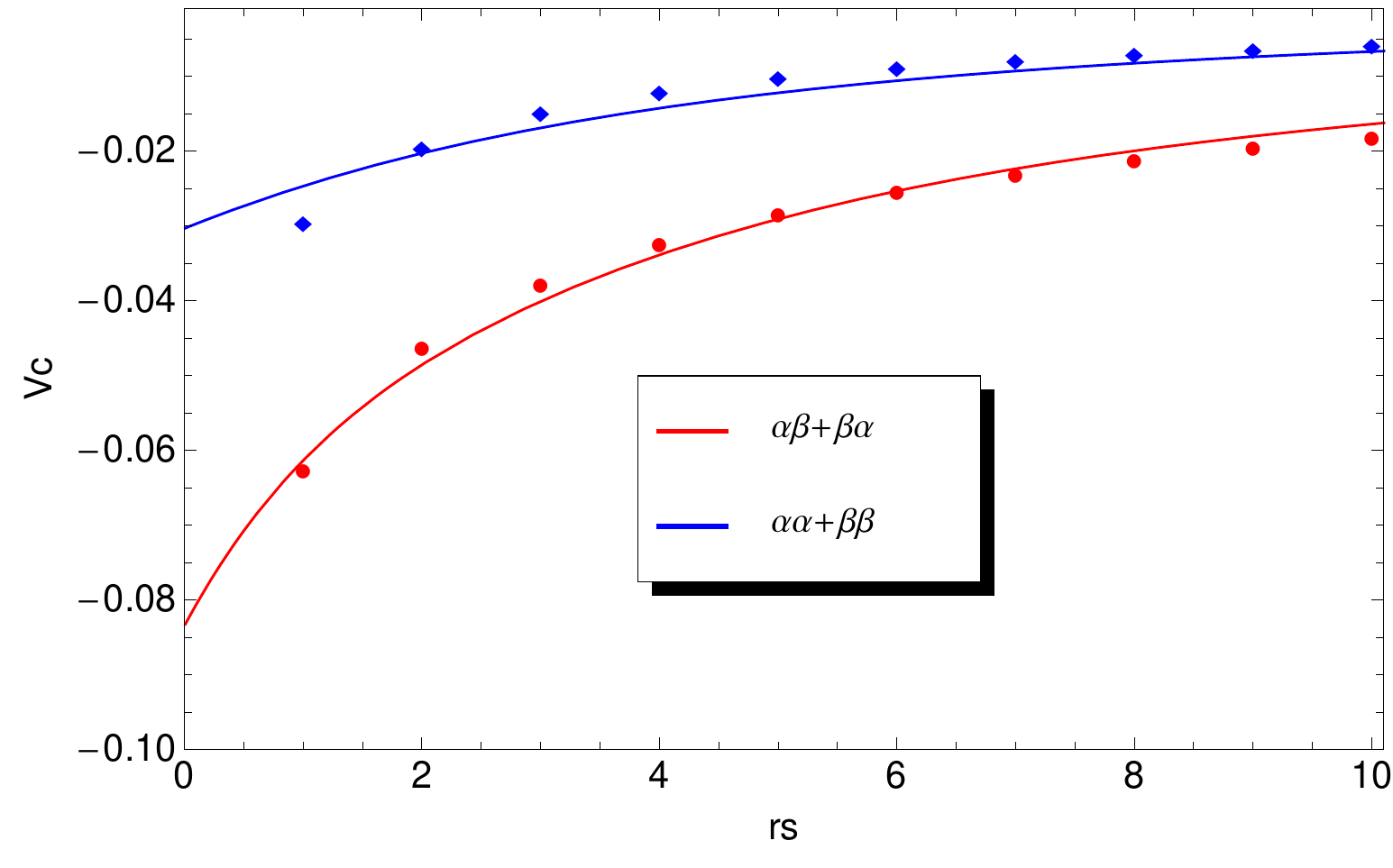}
\caption{Comparison of the approximate
  $V_\mathrm{C}^{\sigma\sigma',\lambda=1}$ energies in the HEG limit with reference
  Monte Carlo data.\cite{gori2000analytic} Solid lines refer
  to Eqs~\ref{vcabl} and~\ref{vcaal}. Circles and diamonds represent reference values.
  The cause of discrepancy at low $\rs$ (high densities) is
  discussed in the main text. Atomic units are used.}
\label{fig:vc}
\end{figure}

Several requisites for the exact exchange-correlation functional were derived using uniform coordinate scaling
technique,\cite{levy1991density,levy1985hellmann} i.e. by applying uniformly scaled density
\begin{equation}
\rho_\kappa(\ra) = \kappa^3 \rho(\kappa \ra).
\end{equation}
These relations are particularly valuable because they hold for arbitrary $N$-electron densities.
A density-scaling identity proved by Levy,\cite{levy1991density}
\begin{equation}
\hcss(\rho;\ra,s) = \lambda^3 h_{\mathrm{C}\lambda'=1}^{\sigma\sigma'}%
(\rho_{1/\lambda}; \lambda\ra,\lambda s),\label{lambda-dependence}
\end{equation}
constrains the set of admissible forms of $\dopp$ and
$\dpar$. \refr{lambda-dependence} implies that
\begin{equation}
d_{\sigma\sigma'}(\rho,|\nabla\rho|,\lambda) = \lambda d_{\sigma\sigma'}%
\left(\frac{\rho}{\lambda^3},\frac{|\nabla\rho|}{\lambda^4},1\right).
\end{equation}
We propose the following simple function which satisfies the scaling condition:
\begin{equation}
d_{\sigma\sigma'} = \frac{F_{\sigma\sigma'}}{\rs^{\sigma\sigma'}} 
+ \frac{G}{\rs} \frac{\nabla\rho \cdot \nabla\rho}{\rho^{8/3}}.
\label{dss-form}
\end{equation}
As~\refr{dss-form} is independent of $\lambda$, the coupling-constant
integration of~\refr{lambda-average} can be done analytically. The values
of  $F_{\alpha\beta}$ and $F_{\alpha\alpha}$ were determined by least-squares fit of the HEG limit of
 $V_\mathrm{C}^{\alpha\beta,\lambda=1}$ and~$V_\mathrm{C}^{\alpha\alpha,\lambda=1}$ to the reference values.\cite{gori2000analytic}
Opposite-spin and parallel-spin components were fit independently. Reference values of 
$V_\mathrm{C}^{\sigma\sigma',\lambda=1}$ for the HEG were obtained by Gori-Giorgi et~al.\cite{gori2000analytic}
 by integrating pair correlation functions from quantum Monte Carlo simulation.\cite{ortiz1999zero} 
 Our estimates of $V_\mathrm{C}^{\alpha\beta,\lambda=1}$
 and~$V_\mathrm{C}^{\alpha\alpha,\lambda=1}$ were optimized to recover
 the reference values for spin-compensated system at metallic
 densities ($\rs = 1, 2, 3, \ldots, 10$). The resulting parameters are $F_{\alpha\beta}=2.1070$ and
 $F_{\alpha\alpha}=2.6422$. The corresponding mean absolute percentage errors
 (MAPE) of opposite-spin and parallel-spin components are $5.0\%$ and
 $12.0\%$, respectively. The MAPE of total
 $V_\mathrm{C}^{\lambda=1}$ is equal to $4.6\%$. See Fig.~\ref{fig:vc}
 for comparison of our fit to the reference values. At high densities ($\rs < 1$) our model
 does not reduce to the accurate correlation functional for the HEG as it
does not account for the logarithmic divergence of the correlation
energy density for $\rs\rightarrow 0$.\cite{gell1957correlation} 
This is, however, a peculiarity of the HEG that
is not present in finite molecular systems.

We supply our SL correlation functional with the DFT-D3
dispersion correction of~\citet{grimme2010consistent}, which contributes
damped terms of the multipole expansion of the dispersion energy.
The adjustment of the SL part to harmonize with the NL
correction is accomplished by optimization of the $G$
parameter, see~\eqref{dss-form}. The
$G$ parameter can be adjusted freely, without
interfering with any of the above-mentioned physical and formal constraints. In particular, it does not alter the first two terms in the short-range
 spatial Taylor expansion of the correlation holes.
 The value of $G$ can be chosen so that the correlation
 contributions described by SL and NL parts do not overlap. As $G\rightarrow 0$,
 our SL correlation model reduces to the correlation of the HEG
 with self-interaction removed from parallel-spin part. 
 Our numerical results show that this leads to
 a systematic overestimation of intermolecular interactions.
 On the other hand, when $G\rightarrow \infty$,
 the SL correlation vanishes, and the SL functional reduces to an exchange-only
 approximation (without adding the dispersion correction). Provided that the
 exchange functional is free from artificial binding,
 the interaction energies should be underestimated in this limit. Between these two limits
 lays the optimal $G$, which corresponds to an
 interaction curve slightly shallower than the real one, for the addition
 of the negative dispersion term should move the interaction energy towards
 the accurate value.

The $G$ parameter of~Eq.~\ref{dss-form} and the two empirical parameters
present in DFT-D3 dispersion correction, $s_{r,6}$ and $s_8$, (see
Eqs. 3 and 4 in Ref.~\citenum{grimme2010consistent}) were
chosen to optimize mean absolute percentage error of binding energies
in S22 set of non-covalently bound
complexes.\cite{jurecka2006benchmark} The numerical optimization has
been carried out with the constraint that the dispersion-free energy
cannot fall below the reference total interaction energy. During the optimization process,
self-consistent KS calculations in aug-cc-pVTZ basis set were
performed using the molecular
structures published in Ref.~\citenum{jurecka2006benchmark}. Reference
interaction energies were taken from
Ref.~\citenum{podeszwa2010improved}. The resulting optimal values are
$G=0.096240$,  $s_{r,6}=1.1882$, and
$s_8=0.65228$. The interaction energies in S22 set are presented in Table~\ref{s22-results}.

 \section{Implementation}
 The expression for the correlation energy is obtained after inserting~\refr{vcabl}
 and~\refr{vcaal} into~\refr{lambda-average} and integrating with
 respect to $\lambda$. Below we present $\ec$ in a form convenient for implementation.
 \begin{align}
   \ec &= \ecab + \ecba + \ecaa + \ecbb, \\ 
   \ecab &= \int_0^1\vcab \diff\lambda = \int \diff^3 \ra \rhoa \pi
   \frac{\Bopp + \Aopp\dopp}{\dopp^3}, \\
   \ecaa &= \int_0^1\vcaa \diff\lambda = \int \diff^3 \ra \rhoa \pi
   \frac{8\Bpar +  4\Apar\dpar}{\dpar^5}, \\
   \Aopp &= \frac{\rhob}{\rsab} \left[ \left(-P_0 + \sum_{k=1}^4 P_k
     (\rsab)^k \right)\exp\left( -P_5 \rsab \right) + P_0 \right] -
   \rhob \label{Aopp-def}\\
   \Bopp &= \frac{\rhob}{(\rsab)^2} \left[ \left(-Q_0 + \sum_{k=1}^5 Q_k (\rsab)^k
     \right) \exp\left(-Q_6 \rsab \right) + Q_0 \right] + \dopp \Aopp
   \label{Bopp-def}\\
   \Apar &= \frac{D_\alpha}{3 \rsaa}\left[ \left( -R_0 +
     \sum_{k=1}^2 R_k (\rsaa)^k \right) \exp\left( -R_3 \rsaa \right)
     + R_0 \right] - \frac{D_\alpha}{3} \label{Apar-def}\\
   \Bpar &= \frac{D_\alpha}{6(\rsaa)^2} \left[ \left( -S_0 +
     \sum_{k=1}^3 S_k (\rsaa)^k \right)\exp\left( -S_4 \rsaa \right) +
     S_0 \right] + \dpar \Apar \label{Bpar-def}
 \end{align}
Note that $\ecab=\ecba$. The formula for $\ecbb$ can be obtained by substitution
of spin indices in $\ecaa$. The values of the numerical constants appearing in Eqs.~\ref{Aopp-def}--\ref{Bpar-def} are
listed in~Table~\ref{numerical-constants}. The following functions:
$D_\alpha$, $\rsaa$, and $\rsab$ are defined in
Eqs~\ref{d-inhom}, \ref{rsaa-def}, \ref{rsab-def}, respectively. The
$d_{\sigma\sigma'}$ function, defined in Eq.~\ref{dss-form}, is
parametrized as follows:
\begin{align}
  \dopp &= \frac{2.1070}{\rs^{\alpha\beta}} 
+ \frac{0.096240}{\rs} \frac{\nabla\rho \cdot \nabla\rho}{\rho^{8/3}}, \\
  \dpar &= \frac{2.6422}{\rs^{\alpha\alpha}}
+ \frac{0.096240}{\rs} \frac{\nabla\rho \cdot \nabla\rho}{\rho^{8/3}}.
\end{align}
The parameters appearing in DFT-D3 correction\cite{grimme2010consistent} are $s_{r,6}=1.1882$ and
$s_8=0.65228$. Fortran code for numerical evaluation of the correlation
energy and its derivatives, together with the corresponding Mathematica\cite{mathematica7} notebook, can be obtained from the
authors by e-mail or from their webpage. The calculations presented in
this work were done using GAMESS program.\cite{schmidt1993general,gordon2005advances}
\begin{table}[p]
  \caption{Ab initio numerical constants appearing in Eqs~\ref{Aopp-def}--\ref{Bpar-def}}
  \label{numerical-constants}
  \begin{tabular}{l|D{.}{.}{10} D{.}{.}{10} D{.}{.}{10} D{.}{.}{10}}
    \hline \hline
    k &     P_k    &   Q_k      & R_k       &   S_k \\
    \hline
    $0$ &  1.696                 &  3.356               &  1.775                &  3.205 \\
    $1$ & -0.2763                & -2.525               &  0.01213              & -1.784 \\
    $2$ & -0.09359               & -0.4500              & -4.743\times 10^{-3}   &  3.613\times 10^{-3} \\
    $3$ &  3.837\times 10^{-3}    & -0.1060              &  0.5566               & -4.743\times 10^{-3} \\
    $4$ & -2.471\times 10^{-3}    &  5.532\times 10^{-4}  &                       &  0.5566 \\
    $5$ &  0.7524                & -2.471\times 10^{-3}  &                       &          \\
    $6$ &                        &  0.7524              &                       &        \\
    \hline \hline
  \end{tabular}
\end{table}

\section{Discussion}
 Similar strategy for designing a correlation functional, i.e., constructing
 a real-space model for a spin-resolved correlation hole, was originally proposed
 by~\citet{rajagopal1978short} with the first application by~\citet{becke1988correlation}, followed by the works of~\citet{proynov1994simple} and~\citet{tsuneda1999new}. A central role in those methods is played by \emph{correlation length}, a function completely determining both short- and long-range behavior of the correlation holes present in those models. Our approach has more degrees of freedom, as the short-range
 behavior of $\hcss$ is decoupled from the choice of the
 $d_{\sigma\sigma'}$ function which controls its decay. This flexibility allows us to adjust $d_{\sigma\sigma'}$ to
 match a specific NL correction without sacrificing the short-range correlation that
 can be accurately represented by an SL functional.

The ultimate goal is to develop a general-purpose functional that not
only yields satisfactory results for the dispersion interactions,
but also performs not worse than the existing approximations in
predicting other properties of chemical interest. To do so, the
inclusion of the NL constituent and the accompanying adjustment of the
SL part should not affect any of the energetically important
constraints already satisfied by the meta-GGA rung
functionals.\cite{perdew2005prescription} Among the formal
constraints, the most fundamental one is the non-positivity condition,
\begin{equation}
 \vcss\left[\rho\right] \le 0, \label{non-positivity}
\end{equation}
which is obeyed by our model for every spin-density. Similarly,
 the scaling conditions formulated by Levy,\cite{levy1991density} e.g.,
\begin{align}
  \lim_{\kappa\rightarrow 0} \frac{\ec\left[ \rho_\kappa
      \right]}{\kappa} &=
  \sum_{\sigma\sigma'}\lim_{\lambda\rightarrow\infty} \label{low-density-limit}
  \vcss\left[\rho\right] > -\infty, \\
  \frac{\partial \vcss\left[ \rho \right]}{\partial \lambda}  &\le 0,
\end{align}
together with the high-density limit of the
correlation functional,\cite{levy1989asymptotic}
\begin{equation}
   \lim_{\kappa\rightarrow\infty} \ec\left[\rho_\kappa\right] >
  -\infty, \label{high-density-limit}
\end{equation}
are satisfied. A failure to satisfy condition~\eqref{high-density-limit}
may contribute to overbinding of
molecules.\cite{levy1991density} 

In addition to conditions~\eqref{non-positivity}--\eqref{high-density-limit}, our approximation satisfies a constraint which
has a direct connection to the prediction of interaction energies. It
was observed by~\citet{kamiya2002density} that if an SL
functional yields nonzero contributions to the correlation in the tail
of the density, then adding an NL correction may lead to a
severe overbinding.\cite{kamiya2002density} In the tail of electronic
density, where the reduced gradient is large, the $d_{\sigma\sigma'}$ function of~\refr{dss-form} goes to infinity,
thus our SL correlation correctly vanishes.

The correlation self-interaction error is corrected using $\tau_\sigma$
variable (the kinetic energy density), see~\refr{baa-inhom}, similarly to other meta-GGA correlation functionals\cite{becke1988correlation,perdew1999accurate,zhao2006design}. As a result, in our
model the parallel-spin correlation vanishes for single-orbital spin-compensated densities, and
the total correlation energy is zero for hydrogen atom.

The above-mentioned constraints are merely formal prerequisites for a high-quality
 approximation to $\ec$. A decent approximate model should also capture the physics of molecular
 systems. Our model reflects the following physical properties:
 \begin{enumerate}
  \item Short-range electronic correlation is modeled by an expression
    borrowed from the homogeneous electron gas, which is also appropriate
    for real
    systems.\cite{burke1994local,burke1998semilocal,henderson2004short,cancio2000exchange}
    (See Eqs~\ref{bab-heg}, \ref{baa-heg}, and~\ref{baa-inhom}.) To
    the best of our knowledge, we present the first beyond-LDA
    functional which incorporates analytic
    representation of the
    short-range correlation function of the HEG developed
    by~\citet{gori2001short}.
  \item Long-range behavior of the correlation hole is governed by
    $d_{\sigma\sigma'}$ function (see Eqs~\ref{opphole-def},
    \ref{parhole-def}, and~\ref{dss-form}), which depends on both
    density and its gradient at a reference point. This function accounts for damping
    effect of density inhomogeneity ($\nabla\rho_\sigma$) on the correlation
    hole. Furthermore, the $d_{\sigma\sigma'}$ function contains a free
    parameter, $G$. It is used in tuning 
    the spatial range separation of the
    correlation hole to properly blend with the long-range correlation correction.
  \item Our model closely approximates
    the exact correlation in the HEG regime at metallic densities. (See also
    the discussion below~\refr{dss-form}.)
 \end{enumerate} 

 To make our concept of stitching SL and NL correlation more transparent,
 we briefly discuss it in the context of range-separated approach
 of~\citet{kohn1998van}. It is possible to solve the dispersion problem
 within DFT by partitioning the interelectron repulsion, $1/r$, into short-range
 part, $\exp\left(-\mu r\right)/r$, and its long-range
 complement.\cite{kohn1998van} ($\mu$ is a constant.) Both short-range exchange and correlation 
 are then treated at (semi)local level, and the
 contributions originating from the long-range interaction are
 approximated by a formula that is consistent with the
 Casimir-Polder expression in the asymptotic region. Our treatment follows the same general
 idea. The difference is as follows: the exponential factor that damps the interelectronic
 interaction, $\exp\left(-\mu r \right)$, is
 replaced by the $\exp\left(- d_{\sigma\sigma'} r \right)$ function of
 Eqs~\ref{opphole-def}--\ref{parhole-def} which damps the short range
 expansion of the approximate correlation holes. Thus, the $\mu$
 constant is generalized into $d_{\sigma\sigma'}$ function, which depends on
 density and its gradient at a reference point.

\section{Numerical results}
Our aim was to validate the correlation functional presented in this
work, preferably without the interference from the errors of
an exchange approximation. Therefore, we decided to perform
calculations using our correlation combined with full HF-like exchange,
and to compare it with other DFAs involving full HF-like
exchange. Although a general-purpose approximation cannot be formed by
combining semilocal DFT correlation with full exact exchange, it is a
demanding and useful test for a correlation functional. If the correlation
functional performs well with large portion of the exact exchange,
 then there is much room for adjusting the exchange part of a global hybrid or
a range-separated hybrid exchange-correlation functional.

All DFT and HF calculations presented below were performed in
aug-cc-pVTZ basis. All energies are obtained from self-consistent calculations. Table~\ref{s22-results} 
contains interaction energies for S22 set of
molecules.\cite{jurecka2006benchmark} The reference energies, $E_\mathrm{ref}$, are taken from
Ref.~\citenum{podeszwa2010improved}. $E_\mathrm{int}$ denotes
interaction energy calculated using the correlation functional described in this
work combined with $100\%$ HF-like exchange and DFT-D3
correction. All energies, as well as mean signed errors (MSE) and mean
unsigned errors (MUE) are given in kcal/mol. Mean absolute percentage
errors (MAPE) are given in percent. Our results (MUE=$0.46$ kcal/mol) compare rather favorably to
the other methods utilizing full HF-like exchange,
VV09\cite{vydrov2010implementation} (MUE=$0.90$ kcal/mol), M06HF\cite{zhao2008m06,goerigk2011thorough}
(MUE=$0.62$ kcal/mol), and M06HF-D3\cite{zhao2008m06,goerigk2011thorough} (MUE=$0.84$ kcal/mol). The
dispersion-free interaction energies are always significantly below the values
that would be obtained if the dispersion term as defined in SAPT was subtracted,
 see supplementary information in~Ref.~\citenum{pernal2009dispersionless} and~Ref.~\citenum{modrzejewski2012dispersion}. This fact suggests that in our
model, at equilibrium distances, a large fraction of the dispersion
interaction is treated as short-range and accounted for by the SL
functional.

We further evaluate the performance of our approximation on the set of systems from nonbonded interaction
 database of Zhao and Truhlar.\cite{zhao2005design,zhao2005benchmark} This
database gathers interacting dimers in subsets according to the dominant 
character of the interaction: dispersion-dominated (WI7/05 and PPS5/05 subsets), 
dipole interaction (DI6/04 subset), hydrogen-bonded (HB6/04), and charge transfer (CT7/04). 
The results are presented in Tables~\ref{wi-results}, \ref{pps-results}, \ref{di-results},
 \ref{hb-results}, and \ref{ct-results}, respectively.
The reference energies ($E_\mathrm{ref}$) are calculated at
 CCSD(T)/mb-aug-cc-pVTZ level, see Ref.~\citenum{pernal2009dispersionless}. We compare
our approximation ($E_\mathrm{int}$) with M06HF functional\cite{zhao2008m06} ($E_\mathrm{M06HF}$) which combines
empirically-parametrized meta-GGA correlation with full HF-like exchange. As expected, our model
predicts interaction energies more accurately in cases where the dispersion interaction
dominates, see Tables~\ref{wi-results} and~\ref{pps-results}. In case of hydrogen bonded complexes,
 Table~\ref{hb-results}, MAPE of either functional is close to $5\%$. Larger errors are present in DI6/04 
and CT7/04 subsets. Although our approximation performs better that M06HF in case of DI6/04 dimers, the 
error is rather large. In this case, as is seen in Table~\ref{di-results}, both functionals underestimate the interaction
strength and their errors are correlated. This fact suggests that the contribution coming from full HF-like
exchange is too repulsive, which cannot be counterbalanced by semilocal DFT correlation. Both functionals display
largest errors in charge-transfer complexes. Our approximation underestimates interaction for every CT complex.
This behavior to a large degree results from huge errors of the HF theory itself, see  $E_\mathrm{HF}$ column in Table~\ref{ct-results}. As explained by~\citet{cohen2012challenges} this problem can be traced to the localization error of the HF theory, which makes electrons
excessively localized on the monomers. This error manifests itself as a concave curve of energy vs. fractional number of electrons, $E(N)$.\cite{cohen2012challenges} It is also known that pure semilocal DFT approximations
give convex $E(N)$.\cite{cohen2012challenges} See Ref.~\citenum{modrzejewski2012dispersion} for the relevant discussion of \ce{NH3\bond{...}ClF} dimer. Therefore, adding some amount of semilocal exchange 
to our approximation should make the $E(N)$ dependence more linear and make the exchange contribution in CT interactions less repulsive, the step which will be undertaken in the future.

\begin{table}[p]
  \caption{Interaction energies in S22 set (kcal/mol).}
\label{s22-results}
{\tiny
\begin{tabular}{lrrr}
\hline \hline
Dimer & $E_\mathrm{ref}$ & $E_\mathrm{int}$ & $E_\mathrm{dispfree}$ \\
\hline
\bf{Hydrogen-bonded} &  & & \\ 
 \ce{(NH3)2}  & -3.145 & -2.75 & -2.17 \\ 
 \ce{(H2O)2}  & -5.004 & -4.86 & -4.41 \\ 
 Formic acid dimer  & -18.751 & -20.17 & -18.75 \\ 
 Formamide dimer  & -16.063 & -16.46 & -14.86 \\ 
 Uracil dimer planar ($C_{2h}$)  & -20.643 & -21.30 & -19.10 \\ 
 2-pyridone $\cdot$ 2-aminopyridine  & -16.938 & -16.33 & -13.67 \\ 
 Adenine $\cdot$ thymine WC  & -16.554 & -16.15 & -13.24 \\ 
 \hline
 MSE &  -0.13 &  & \\
 MUE &   0.58 &  & \\ 
 MAPE &  5.0 &  & \\
                            & & & \\
\bf{Predominant dispersion} & & &  \\
 \ce{(CH4)2}  & -0.529 & -0.60 & 0.14 \\ 
 \ce{(C2H4)2}  & -1.482 & -1.52 & -0.15 \\ 
 Benzene $\cdot$ \ce{CH4}  & -1.448 & -1.45 & 0.10 \\ 
 Benzene dimer parallel-displaced  ($C_{2h}$)  & -2.655 & -1.76 & 2.55 \\ 
 Pyrazine dimer  & -4.256 & -3.36 & 0.99 \\ 
 Uracil dimer stacked ($C_2$)  & -9.783 & -9.97 & -3.63 \\ 
 Indole $\cdot$ benzene stacked & -4.523 & -3.25 & 2.73 \\ 
 Adenine $\cdot$ thymine stacked  & -11.857 & -11.63 & -3.10 \\ 
 \hline
 MSE &  0.37 &  & \\
 MUE &  0.45 &  & \\ 
 MAPE & 13 &  & \\
                         &  &  & \\
 {\bf Mixed interaction} &  &  &  \\ 
 Ethene $\cdot$ ethyne  & -1.503 & -1.63 & -0.91 \\ 
 Benzene \ce{H2O}  & -3.280 & -3.78 & -2.19 \\ 
 Benzene \ce{NH3}  & -2.319 & -2.55 & -0.92 \\ 
 Benzene \ce{HCN}  & -4.540 & -5.68 & -4.02 \\ 
 Benzene dimer T-shaped ($C_{2v}$)  & -2.717 & -2.72 & -0.18 \\ 
 Indole $\cdot$ benzene T-shaped  & -5.627 & -5.89 & -2.45 \\ 
 Phenol dimer  & -7.097 & -6.87 & -3.96 \\ 
 \hline
 MSE & -0.29 &  & \\
 MUE &  0.36 &  & \\ 
 MAPE & 9.5 &  & \\
 & & & \\
 MSE (total) &  0.002 &  & \\
 MUE (total) &  0.46 &  & \\ 
 MAPE (total) &  9.3 &  & \\
 \hline \hline
\end{tabular}
}
\end{table}

\begin{table}[p]
  \caption{Interaction energies in WI7/05 set (kcal/mol).}
\label{wi-results}
\begin{tabular}{lrrr}
\hline \hline
Dimer  & $E_\mathrm{ref}$ & $E_\mathrm{int}$ & $E_\mathrm{M06HF}$ \\
\hline
\ce{He\bond{...}Ne} & -0.041 & -0.037 & -0.13 \\ 
\ce{He\bond{...}Ar} & -0.058 & -0.045 & -0.085 \\ 
\ce{Ne\bond{...}Ne} & -0.086 & -0.064 & -0.13 \\ 
\ce{Ne\bond{...}Ar} & -0.13 & -0.07 & -0.15 \\ 
\ce{CH4\bond{...}Ne} & -0.18 & -0.18 & -0.20 \\ 
\ce{C6H6\bond{...}Ne} & -0.41 & -0.53 & -0.66 \\ 
\ce{CH4\bond{...}CH4} & -0.53 & -0.59 & -0.12 \\ 
\hline
MSE                   &       &  -0.01     &  -0.006     \\
MUE                   &       &   0.04     &  0.12       \\
MAPE                  &       &   21       &  68           \\
\hline \hline
\end{tabular}
\end{table}

\begin{table}[p]
  \caption{Interaction energies in PPS5/05 set (kcal/mol).}
\label{pps-results}
\begin{tabular}{lrrr}
\hline \hline
Dimer  & $E_\mathrm{ref}$ & $E_\mathrm{int}$ & $E_\mathrm{M06HF}$ \\
\hline
\ce{(C2H2)2}           & -1.36 & -1.47 & -1.06 \\ 
\ce{(C2H4)2}           & -1.44 & -1.52 & -0.95 \\ 
Sandwich \ce{(C6H6)2}  & -1.65 & -0.95 & 0.48 \\ 
T-shaped \ce{(C6H6)2}  & -2.63 & -2.78 & -1.95 \\ 
Displaced \ce{(C6H6)2} & -2.59 & -2.06 & -0.94 \\ 
\hline
MSE                    &       & 0.18      & 1.0  \\
MUE                    &       & 0.31      & 1.0  \\
MAPE                   &       & 16        & 55   \\
\hline \hline
\end{tabular}
\end{table}

\begin{table}[p]
  \caption{Interaction energies in DI6/04 set (kcal/mol).}
\label{di-results}
\begin{tabular}{lrrr}
\hline \hline
Dimer  & $E_\mathrm{ref}$ & $E_\mathrm{int}$ & $E_\mathrm{M06HF}$ \\
\hline
\ce{H2S\bond{...}H2S}   & -1.62 & -1.20 & -0.82 \\ 
\ce{HCl\bond{...}HCl}   & -1.91 & -1.40 & -0.99 \\ 
\ce{HCl\bond{...}H2S}   & -3.26 & -2.74 & -2.48 \\ 
\ce{CH3Cl\bond{...}HCl} & -3.39 & -2.77 & -2.72 \\ 
\ce{CH3SH\bond{...}HCN} & -3.58 & -3.70 & -3.50 \\ 
\ce{CH3SH\bond{...}HCl} & -4.74 & -4.13 & -4.27 \\ 
\hline
MSE                     &       &   0.43    &   0.62      \\
MUE                     &       &   0.47    &   0.62      \\
MAPE                    &       &   17      &     26        \\
\hline \hline
\end{tabular}
\end{table}

\begin{table}[p]
\caption{Interaction energies in HB6/04 set (kcal/mol).}
\label{hb-results}
\begin{tabular}{lrrr}
\hline \hline
Dimer  & $E_\mathrm{ref}$ & $E_\mathrm{int}$ & $E_\mathrm{M06HF}$ \\
\hline
\ce{NH3\bond{...}NH3}  & -3.09  & -2.82 & -2.53 \\    
\ce{HF\bond{...}HF}    & -4.49  & -4.63 & -4.27 \\    
\ce{H2O\bond{...}H2O}  & -4.91  & -4.90 & -4.72 \\    
\ce{NH3\bond{...}H2O}  & -6.38  & -6.28 & -6.35 \\    
\ce{(HCONH2)2}         & -15.41 & -16.39 & -15.72 \\ 
\ce{(HCOOH)2}          & -17.60 & -19.57 & -19.33 \\ 
\hline
MSE                    &        &  -0.45     &  0.28           \\
MUE                    &        &  0.58      &  0.30           \\
MAPE                   &        &  5.2       &  4.6            \\
\hline \hline
\end{tabular}
\end{table}

\begin{table}[p]
\caption{Interaction energies CT7/04 (kcal/mol).}
\label{ct-results}
\begin{tabular}{lrrrr}
\hline \hline
Dimer  & $E_\mathrm{ref}$ & $E_\mathrm{int}$ & $E_\mathrm{M06HF}$ & $E_\mathrm{HF}$ \\
\hline
\ce{C2H4\bond{...}F2}  & -1.06 & -0.33 & -0.67    & 0.71 \\  
\ce{NH3\bond{...}F2}   & -1.80 & -0.74 & -0.90    & 0.19 \\  
\ce{C2H2\bond{...}ClF} & -3.79 & -2.98 & -4.18    & -0.16 \\ 
\ce{HCN\bond{...}ClF}  & -4.80 & -3.70 & -4.02    & -2.10 \\ 
\ce{NH3\bond{...}Cl2}  & -4.85 & -3.50 & -4.00    & -1.12 \\ 
\ce{H2O\bond{...}ClF}  & -5.20 & -4.69 & -5.26    & -2.91 \\ 
\ce{NH3\bond{...}ClF}  & -11.17 & -10.53 & -11.92 & -5.49 \\ 
\hline
MSE                    &        &   0.89    &   0.24   &     \\
MUE                    &        &   0.89    &   0.59   &     \\
MAPE                   &        &   31      &   20     &     \\
\hline \hline
\end{tabular}
\end{table}

\section{Conclusions}
This paper presents  a novel form of an SL correlation functional belonging to the meta-GGA rung that may be combined in an optimal way with the dispersion interaction component, either in the DFT+D manner or by incorporating a nonlocal potential. The important feature is that it is  based on the first principles, in the form of a number of physical constraints imposed during the derivation. With minimal empiricism, our approximation is adjusted to a desired long-range dispersion correction by optimizing only a single empirical parameter. The parameter has a clear physical meaning: it governs the decay of the approximate correlation hole. Consequently, the correlation hole vanishes exponentially at large inter-electronic distances, which prevents double counting of the electron correlation effect that is already included when adding the long-range dispersion correction. An important and unique facet of our functional is that the adjustment of the empirical range-separation parameter has \emph{not} relaxed any of the physical constraints on which our model is based. The electron correlation is approximated by utilizing several numerical and analytical results of the HEG model. Most importantly, the HEG approximation to the short-range part of the correlation hole is rigorously  conserved for arbitrary systems (only the self-interaction pertinent to the HEG model is removed from the parallel-spin correlation hole).

While our new correlation functional can be combined with any of non-local dispersion models, for preliminary calculations of this paper, we employed the atom pairwise additive DFT-D3 dispersion correction. Given the fact that  our correlation functional is combined with 100\% HF exchange -- far from an optimal choice in general case -- the numerical results are very encouraging. For the interaction energies of hydrogen-bonded complexes, the accuracy is on a par with that obtained with the M06HF functional, which is a highly parametrized empirical approximation containing full HF exchange. For dispersion-dominated complexes, the predictions of our model compare favorably with VV09 and M06HF. The results in the subsets of dipole-interaction and charge-transfer complexes are less satisfactory, which is easily explained by the inadequacy of the full HF-like exchange component: indeed, the signed errors correlate with the  signed errors of the HF method. Obviously, much improvement may be expected when a more appropriate exchange part will be incorporated.  Development of an optimal range-separated hybrid exchange approximation, appropriate for our new correlation functional, as well as implementation of non-local van der Waals correlation functionals are underway in our laboratory.

\section{Acknowledgments}
This work was supported by the Polish Ministry of Science and Higher
Education, Grant~N~N204~248440, and by the National Science Foundation
(US), Grant No. CHE-1152474.
\bibliography{biblio}
\end{document}